\documentstyle[floats,prl,aps]{revtex}

\input epsf

\begin{document}

\preprint{CfPA-TH-96}
\title{Observational Constraints on Open Inflation Models}
\author{Martin White${}^1$ and Joseph Silk${}^2$}
\address{${}^1$Enrico Fermi Institute, University of Chicago\\
Chicago, IL 60637 \\
${}^2$Center for Particle Astrophysics and Departments of Astronomy and
Physics\\ University of California, Berkeley CA 94720}

\twocolumn[
\maketitle
\widetext
% \centerline{CfPA-96-TH-16}

\begin{abstract}
We discuss observational constraints on models of open inflation.
Current data from large-scale structure and the cosmic microwave background
prefer models with blue spectra and/or $\Omega_0\ge0.3$--0.5.
Models with minimal anisotropy at large angles are strongly preferred.
\end{abstract}
\hskip 0.5truecm
]
\narrowtext

In this Letter we examine the parameter space allowed by models of structure
formation in a universe of subcritical density in which the main constituent
is cold dark matter (CDM): $\Omega_{\rm CDM}\equiv\Omega_0<1$.
In spatially flat models with a cosmological constant, the initial fluctuation
spectrum is well-defined and such models have received considerable attention
\cite{lambda,LLVW}.
However there is rather more structure in open inflationary models
\cite{open,BucTur,openintro}, and we believe it is timely to consider
the constraints that arise from the formation of large-scale structure
and the observed anisotropies in the cosmic microwave background (CMB).
We apply these constraints to provide guidance for constructing open
inflationary models by emphasizing the types of anisotropy spectra which are
preferred by the data.  None of the models currently  discussed in the
literature have the features necessary for a low-$\Omega_0$ universe with a
Hubble constant $>60$--$65\ {\rm km}\,{\rm s}^{-1}\,{\rm Mpc}^{-1}$.

The closest progenitor of this work is that of \cite{LLRV}, who looked at the
large-scale structure constraints for open models normalized to the {\sl COBE}
2-year data.  An in-depth treatment of current open models with scale-invariant
spectra below the curvature scale has recently appeared in \cite{Goretal},
although their emphasis  differs from the present one.
No previous discussions have focussed on the full range of allowable parameter
space in CDM models, including both large-scale structure and CMB constraints.

Both observational evidence and considerations of inflation
\cite{lambda,open,BucTur}
motivate a model possessing adiabatic fluctuations and CDM, with, if the
universe is indeed at subcritical density, preferably $\Omega_0\in[0.1,0.5]$.
Direct measures of the Hubble constant appear to be approaching a consensus
that $h=0.65\pm 0.1$, where $H_0=100h\ {\rm km}\,{\rm s}^{-1}\,{\rm Mpc}^{-1}$.
We will therefore focus attention on this range of $(\Omega_0,h)$ parameter
space.
All of our models will be normalized to the 4-year {\sl COBE}-DMR data
\cite{Ben} as described in \cite{BunWhi}.
Specifically, we take the amplitude of the density fluctuations at
horizon-crossing to be
\begin{eqnarray}
\delta_{\rm H}(n,\Omega_0)  & = & 
        1.95\times 10^{-5}\;
	\Omega_0^{-0.35-0.19\ln\Omega_0-0.17\widetilde{n}} \\ \nonumber
&&      \exp \left[ -\widetilde{n} -0.14\widetilde{n}^2 \right] \,.
\label{eqn:dhopen}
\end{eqnarray}
where $\widetilde{n}=n-1$ and $\delta_{\rm H}$ is related to the matter power
spectrum today by
\begin{eqnarray}
\Delta^2(k) &\equiv& {k^3 P(k)\over 2\pi^2} \\
&\equiv&
  \left( {k\over H_0} \right)^{3+n}\ \delta_{\rm H}^2\ T^2(k) \, ,
\end{eqnarray}
and $T(k)$ is the matter transfer function which depends on the cosmological
parameters $\Omega_0$ and $h$.
This normalization assumes that the fluctuations in the gravitational potential
are a power law, $k^{n-1}$, in the eigenvalue $k$ of the Laplacian and that
only sub-curvature scalar perturbations give rise to the CMB anisotropy.  We
will return to these points in detail later.

Restricting ourselves to open inflationary models, our parameter space consists
of $\Omega_0<1$, $n$, $h$ and $\Omega_{\rm B}h^2$.
There are a variety of observational tests which any model of structure
formation must pass (see e.g.~\cite{tests,LLRV,LLVW}).
For our models, the most constraining tests, based on the well defined linear
theory predictions, are the shape of the CDM power spectrum $\Delta^2$
\cite{PD} and the cluster abundance \cite{VL}.
For the former, we perform a $\chi^2$ fit directly to the data for $\Delta^2$
as tabulated in \cite{PD}, excluding the last 4 points and allowing the overall
normalization to float.  The latter constraint can be expressed as a limit on
\begin{equation}
\sigma^2_8 = \int {{\rm d}k\over k}\ \Delta^2(k)\ W^2(kR) 
\end{equation}
where $R=8h^{-1}{\rm Mpc}$, which probes scales $k\sim 0.2h\,{\rm Mpc}^{-1}$.
The amount of small-scale power required to fit the cluster abundance in these
models is \cite{VL},
\begin{equation}
  \sigma_8 = 0.6\ \Omega_0^{-0.36-0.31\Omega_0+0.28\Omega_0^2}
\end{equation}
where the errors are $+32\%$ $\times\Omega_0^{0.17\log_{10}\Omega_0}$
and $-24\%$ $\times\Omega_0^{0.17\log_{10}\Omega_0}$ at 95\%CL.
To this we add in quadrature a 2$\sigma$ error from the {\sl COBE}
normalization of 20\%.
We have also checked that the abundances of high redshift massive objects
(e.g.~quasars and damped Ly$\alpha$ systems) are compatible with the
observations for all of these models.  Further, we require that our models are
at least $12\,$Gyr old \cite{CKKD}, which provides an upper limit to the value
of $\Omega_0$ which we can consider for any fixed Hubble constant.
Since we will be concerned here mostly with the {\it lower} limits to
$\Omega_0$, the precise age constraint that we impose will not be important.

\begin{figure}[t]
\begin{center}
\leavevmode
\epsfxsize=3.5in \epsfbox{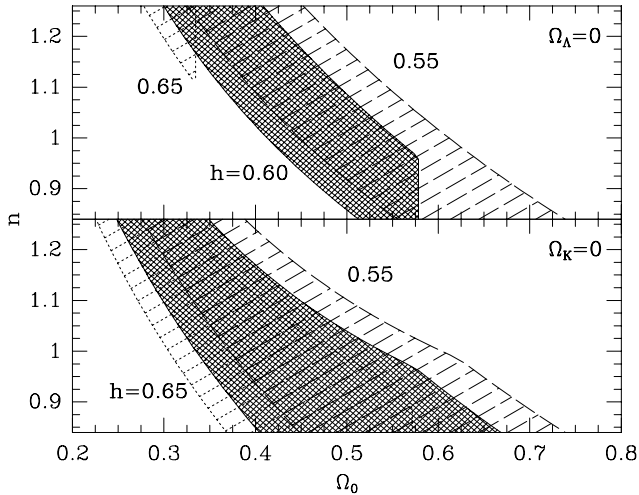}
\end{center}
\caption{\noindent Allowed regions: The values of $\Omega_0$ and $n$, for
fixed $h$, for which CDM models do not violate any of our large-scale structure
constraints at 95\%CL.  Upper limits on $\Omega_0$ come from requiring
$t_0\ge12\,$Gyr, other constraints are discussed in the text.
We have assumed $\Omega_{\rm B}h^2=0.015$ and minimal large-angle CMB
anisotropies.}
\label{fig:lss}
\end{figure}

We show in Fig.~\ref{fig:lss} the  region of parameter space allowed after
applying these constraints for a variety of slices with $h=$constant.
We have fixed the baryon content of the universe at this stage to
$\Omega_{\rm B}h^2=0.015$.
Our conclusions do not depend strongly on this assumption for the range of
$\Omega_0$ we consider.  Lowering $\Omega_{\rm B}h^2$ allows slightly higher
$\Omega_0$ for fixed $h$ while increasing it allows slightly lower $\Omega_0$.

Let us concentrate on the open models (top panel); we present the flat
models (lower panel) merely for comparison.
Notice that very low-$\Omega_0$ models are only allowed if the primordial
fluctuation spectrum is ``blue'' ($n>1$).  Such blue spectra arise in
``flat'' inflation \cite{blue} mostly in hybrid or two-field models where
inflation ends not by a field rolling down an ever steepening potential but
through an instability.  To the best of our knowledge no models of open
inflation have been constructed with this property, though models with
``blue'' specta may be more natural in open inflation \cite{BucTur}.

The problem that arises in fitting the data to these models is due to the low
{\sl COBE} normalization \cite{WhiSco,BunWhi}.
In each case the upper limit on $\Omega_0$ for fixed $n$ comes from the shape
of the CDM power spectrum, while the lower limit comes from the cluster
abundance.  This latter constraint scales roughly as $\Omega^{-0.4}$ while the
{\sl COBE} normalization at large scales is almost flat with $\Omega_0$.
At fixed $h$, lowering $\Omega_0$ changes the shape of the CDM transfer
function so as to reduce small-scale power.
The lowest $\Omega_0$ allowed occurs for the largest ratio of small- to
large-scale power, i.e.~when we tilt the primoridal power spectrum to $n>1$
or increase the Hubble constant.
Less tilt, or lower $\Omega_0$, could be accomodated if $\sigma_8$ inferred
from the cluster abundances were lowered.  However we have already adopted
conservative uncertainties on this quantity.

In producing Fig.~\ref{fig:lss}, we have normalized the models to the
{\sl COBE} data assuming that the primordial spectrum of curvature
perturbations is a power-law in the eigenvalue of the Laplacian.  Ignoring
a possible ``running'' of the spectral index, this is reasonable on scales
much smaller than the curvature scale.  However open inflation models can
predict departures from simple power-law behaviour near the curvature scale.
These departures only affect the lowest multipoles of the CMB anisotropy
spectrum, and hence the {\sl COBE} normalization.
For current models, departures from a power-law on {\sl COBE} scales do not
affect the normalization significantly \cite{YamBun}, although models that
decreased the anisotropy at fixed $\delta_{\rm H}$ would provide a better fit
to the data.
More importantly, we have not included any contribution from ``super-curvature
modes'' \cite{supercurv}, ``bubble wall modes'' \cite{bubblewall} or tensor
anisotropies.
Since the {\sl COBE} measurement of the temperature anisotropies is fixed,
any of these ``extra'' contributions to the anisotropy would {\it lower\/}
$\delta_{\rm H}$, worsening the agreement with observations.
For example, decreasing $\delta_{\rm H}$ by a factor of 1.5 reduces the allowed
region to a thin band, $\sim1/5$ of the width, along the upper right of the
regions shown in Fig.~\ref{fig:lss}.
Reducing $\delta_{\rm H}$ by a factor of 2 causes the allowed regions shown
in Fig.~\ref{fig:lss} to disappear entirely (a small region persists at high
$\Omega_0$ and $n\sim0.7$--0.8, which is off our plot).

Thus the challenge to model builders interested in constructing open universe
inflationary models is to provide ``blue'' spectra with a minimal amount
of large-angle CMB anisotropy.  If we imagine that obtaining very ``blue''
spectra ($n\ge 1.1$--$1.2$) is as difficult in the open models as it is in
the flat models (see e.g.~\cite{blue}),
then open models will require a relatively low Hubble constant and a relatively
high density: $\Omega_0\ge 0.4$--$0.5$.
If the models {\it are} significantly ``blue'', then they will not look like
the usual scale-invariant CDM models parameterized by $\Gamma\simeq\Omega_0h$.
Also should the models have power-law spectra which stay ``blue'' to very small
scales, limits from the production of spectral distortions \cite{mudist} and
primordial black holes \cite{bh} need to be considered.

\begin{figure*}[t]
\begin{center}
\leavevmode
\epsfxsize=6.5in \epsfbox{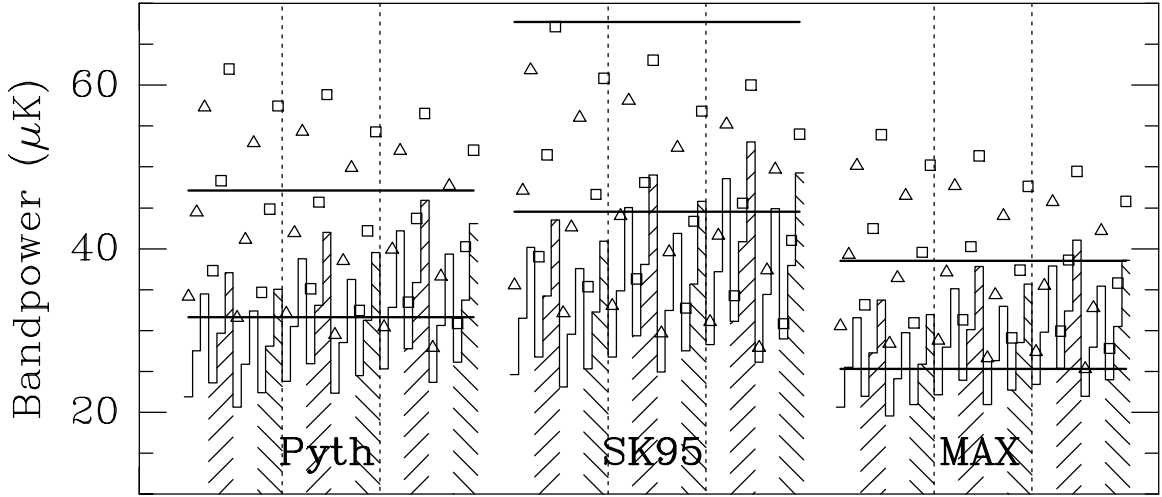}
\end{center}
\caption{\noindent CMB predictions: The bandpower predicted, as a function
of ``model number'', for the Python, Saskatoon and MAX experiments, with
95\%CL measurements (solid, horizontal lines) {\it excluding\/} calibration
uncertainty (Py: 20\%, SK: 14\%, MAX: 10\%) and 10\% {\sl COBE} normalization
uncertainty.  The histogram represents open models.
Unshaded models have $\Omega_{\rm B}h^2=0.01$, shaded models have
$\Omega_{\rm B}h^2=0.02$.  Within each group $n=0.8$, $1$, $1.2$.
Shading running bottom-left to top-right indicates $h=0.6$, top-left to
bottom-right $h=0.7$, with the unshaded predictions having the same
$h$ as the shaded prediction to their right.  Vertical dotted lines separate
$\Omega_0=0.3$, $0.4$ and $0.5$ (left to right).  The open triangles and
squares are the predictions for $\Omega_\Lambda=1-\Omega_0$ for comparison.}
\label{fig:cmb}
\end{figure*}

Another strong constraint on open models comes from small angular scale
measurements of the CMB anisotropy.
It has long been realized  \cite{DZS,WilSil,modern} that the observation
of any feature in the CMB angular power spectrum would allow one to perform
the classical ``angular diameter distance'' cosmological test of the
(spatial) curvature of the universe (e.g.~\cite{Pee}).
In an open universe, the peaks in the angular power spectrum of a CDM model
shift to smaller angular scale.  Thus experiments which probe angular scales
around $0.5^{\circ}$ will observe more power in a model without spatial
curvature than with negative spatial curvature.
To quantify this we have chosen 3 representative experiments which probe
scales near the first peak in a flat model: Python \cite{python},
Saskatoon \cite{sask} and MAX \cite{max}.
For Python we have taken the highest $\ell$ bandpower, for Saskatoon the 3rd
of 5 bandpowers and for MAX the ``combined'' analysis.
We show in Fig.~\ref{fig:cmb} the predicted bandpower (also known as
$Q_{\rm flat}$, see \cite{bandpower}), or level of fluctuation, for 32
models with $\Omega_0=0.3$, 0.4, 0.5 and $n=0.8$, 1.0, 1.2.
Since the CMB anisotropies are sensitive to $h$ and $\Omega_{\rm B}h^2$ we
have computed them for $h=0.6$, 0.7 and $\Omega_{\rm B}h^2=0.01$, 0.02.
Interpolation to other values near these is stable.
These predictions were computed by numerical evolution of the coupled
Einstein, Boltzmann and fluid equations as discussed in \cite{HSSW}, using
the window functions provided by the authors of \cite{python,sask,max}.
We have not included late reionization, which would lower the predictions
by $\exp(-\tau)$ where $\tau$ is the optical depth to Thomson scattering
between the redshift of reionization and today.
This redshift is very uncertain, our best guess (see e.g.~\cite{reion}) puts
$\tau$ in the range $\sim1\%$ to 30\%.

The trends in Fig.~\ref{fig:cmb} are easy to understand.  Higher $n$ means
more fluctuation power at $0.5^{\circ}$ since the models are normalized to
{\sl COBE} on large scales.
Similarly, higher $\Omega_{\rm B}h^2$ means more power on small scales since it
enhances the amplitude of the acoustic oscillations in the baryon--photon
plasma before recombination.  Lowering $h$ shifts matter--radiation equality
closer to last-scattering, thus enhancing the acoustic oscillations due to the
decay of the potentials.
For the open models, lowering $\Omega_0$ enhances the peak height (due to the
shift of equality), but it also shifts the peak to smaller angular scales
(out of the experimental window) and increases the anisotropy at large-scales
(which affects the {\sl COBE} normalization).
The net effect is to {\it lower\/} the predicted power at $0.5^{\circ}$.
For the $\Omega_\Lambda$ models the enhanced peak height serves to raise the
predicted power at $0.5^{\circ}$.

Providing enough power at small (now angular) scales is confirmed to be
a problem for open models with $\Omega_0\ll 1$.
We conclude that those models best able to fit the data have $n>1$,
$\Omega_0\approx0.5$, a relatively low Hubble constant and a high
$\Omega_{\rm B}h^2$. 
Any ``extra'' contribution to the large-angle anisotropy over what we have
assumed, such as super-curvature modes, bubble wall fluctuations or
gravitational waves, would worsen the situation.

In summary, we have compared models of open inflation with the available data
on the shape of the CDM power spectrum, the abundance of rich clusters and
small scale CMB anisotropies.
The {\sl COBE}-DMR normalized models prefer a relatively high $\Omega_0\ge0.3$
and/or ``blue'' spectra $n\ge 1$.
Models in which ``extra'' CMB anisotropies are present are very strongly
constrained: enough extra anisotropy to lower $\delta_{\rm H}$ by a factor
of 2 leaves no allowed region on Fig.~\ref{fig:lss}.
The constraints become stronger should the universe contain some hot as well
as cold dark matter, which reduces the amount of small-scale power
(e.g.~\cite{Primack}).
A similar reduction in small-scale power would be obtained by a large increase
in $\Omega_{\rm B}h^2$ (e.g.~\cite{BurTyt}), especially at low $\Omega_0$.
Alternatively if the $\sigma_8$ inferred from cluster abundance measures is
lowered then the allowed region would be increased.

\bigskip
% \acknowledgments  
We acknowledge useful conversations with Joanne Cohn and Andrew Liddle.


\begin{thebibliography}{99}
\bibitem{lambda}
D. Scott, J. Silk, M. White, Science, 268, 829 (1995), astro-ph/9505015;
L.M. Krauss, M.S. Turner, Gen. Rel. \& Grav., 27, 1137 (1995), astro-ph/9504003;
J.P. Ostriker, P.J. Steinhardt, Nature, 377, 600 (1995), astro-ph/9505066
\bibitem{LLVW}
A.R. Liddle, D.H. Lyth, P.T.P. Viana, M. White, MNRAS, 282, 281 (1996),
  astro-ph/9512102
\bibitem{open}
D.H. Lyth, E.D. Stewart, Phys. Lett. B252, 336 (1990);
M. Sasaki, T. Tanaka, K. Yamamoto, J. Yokoyama,
Phys. Lett. {\bf B317} 510 (1993);
B. Ratra, P.J.E. Peebles, Astrophys. J., 432, L5 (1994);
M. Bucher, A.S. Goldhaber, N. Turok, Phys. Rev. {\bf D52}, 3314 (1995),
  hep-ph/9411206;
A.D. Linde, Phys. Lett. {\bf B351}, 99 (1995), hep-ph/9503097;
A.D. Linde, A. Mezhlumian, Phys. Rev. {\bf D52}, 6789 (1995), astro-ph/9506017;
A.M. Green, A.R. Liddle, Phys. Rev. {\bf D}, submitted, astro-ph/9607166
\bibitem{BucTur}
M. Bucher, N. Turok, Phys. Rev. {\bf D52}, 5538 (1995), hep-ph/9503393
\bibitem{openintro}
J.D. Cohn, in Proceedings of the XXXIth Rencontres de Moriond,
{\it Microwave Background Anisotropies}, astro-ph/9606052
\bibitem{LLRV}
A.R. Liddle, D.H. Lyth, D. Roberts, P.T.P. Viana, MNRAS, 278, 644 (1996),
  astro-ph/9506091.
\bibitem{Goretal}
K. Gorski, B. Ratra, R. Stompor, N. Sugiyama, A.J. Banday, astro-ph/9608054
\bibitem{Ben}
C.L. Bennett, et al., Astrophys. J., 454, L1 (1996), astro-ph/9601067
\bibitem{BunWhi}
E.F. Bunn, M. White, Astrophys. J., submitted, astro-ph/9607060
\bibitem{tests}
M. White, P.T.P. Viana, A.R. Liddle, D. Scott, MNRAS, in press,
  astro-ph/9605057;
A.R. Liddle, D.H. Lyth, R.K. Schaefer, Q. Shafi, P.T.P. Viana,
  MNRAS, 281, 531 (1996), astro-ph/9511057;
\bibitem{PD}
J.A. Peacock, S.J. Dodds, MNRAS, 267, 1020 (1994), astro-ph/9311057.
\bibitem{VL}
P.T.P. Viana, A.R. Liddle, MNRAS, 281, 323 (1996), astro-ph/9511007.
\bibitem{CKKD}
B. Chaboyer, P.J. Kernan, L.M. Krauss, P. Demarque, astro-ph/9509115.
\bibitem{blue}
A.R. Liddle, D.H. Lyth, MNRAS, 265, 379 (1993);
A. Linde, Phys. Rev., D49, 748 (1994);
S. Mollerach, S. Matarrese, F. Lucchin, Phys. Rev., D50, 4835 (1994),
  astro-ph/9309054;
E.J. Copeland, A.R. Liddle, D.H. Lyth, E.D. Stewart, D. Wands,
  Phys. Rev., D49, 6410 (1994), astro-ph/9401011;
J. Garcia-Bellido, D. Wands, astro-ph/9606047
\bibitem{WhiSco}
M. White, D. Scott, Comments Astrophys., 18, 289 (1996), astro-ph/9601170.
\bibitem{YamBun}
K. Yamamoto, E.F. Bunn, Astrophys. J., 464, 8 (1996), astro-ph/9508090;
\bibitem{supercurv}
M. Sasaki, T. Tanaka, K. Yamamoto, Phys. Rev. {\bf D51}, 2979 (1995);
K. Yamamoto, E.F. Bunn, astro-ph/9508090;
M. Sasaki, T. Tanaka, astro-ph/9605104.
\bibitem{bubblewall}
A. Linde, A. Mezhlumian, Phys. Rev. {\bf D52}, 6789 (1995), astro-ph/9506017;
T. Hamazaki, M. Sasaki, T. Tanaka, K. Yamamoto,
  Phys. Rev. {\bf D53} 2045 (1996);
J. Garriga, gr-qc/9602025;
J. Garcia-Bellido, astro-ph/9510029.
\bibitem{mudist}
W. Hu, D. Scott, J. Silk, Astrophys. J., 430, L5 (1994), astro-ph/9402045.
\bibitem{bh}
B.J. Carr, J.H. Gilbert, J.E. Lidsey, Phys. Rev. {\bf D50}, 4853 (1994),
  astro-ph/9405027.
\bibitem{DZS}
A.G. Doroshkevich, Ya$\,$B Zel'dovich, R.A. Sunyaev,
Sov. Astron., 22, 523 (1978).
\bibitem{WilSil}
M.S. Wilson and J. Silk, Astrophys. J., 243, 14 (1981).
\bibitem{modern}
N. Sugiyama, N. Gouda, Prog. Theor. Phys., 88, 803 (1992);
M. Kamionkowski, D. Spergel, N. Sugiyama, Astrophys. J., 426, L57 (1994),
  astro-ph/9401003.
\bibitem{Pee}
P.J.E. Peebles, ``Principles of Physical Cosmology'', (Princeton, NJ) 1993.
\bibitem{python}
S.R. Platt, J. Kovac, M. Dragovan, J.B. Peterson, J.E. Ruhl, Astrophys. J.,
in press, astro-ph/9606175
\bibitem{sask}
C.B. Netterfield, M.J. Devlin, N. Jarosik, L. Page, E.J. Wollack,
  Astrophys. J., in press, astro-ph/9601197
\bibitem{max}
S. Tanaka, et al., in preparation (1996).
\bibitem{bandpower}
M. White, D. Scott, in {\it CMB Anisotropies 2 Years After COBE:
  Observations, Theory and the Future}, ed.~L.~Krauss, p.~254,
  astro-ph/9406060.
\bibitem{HSSW}
W. Hu, D. Scott, N. Sugiyama, M. White, Phys. Rev. {\bf D52}, 5498 (1995),
  astro-ph/9505043.
\bibitem{reion}
H.M.P. Couchman, MNRAS {\bf 214}, 137 (1985).
H.M.P. Couchman, M.J. Rees, MNRAS {\bf 221}, 53 (1986);
M. Fukugita, M. Kawasaki, MNRAS {\bf 269}, 563 (1994);
S. Sasaki, F. Takahara, Y. Suto, Prog. Theor. Phys. {\bf 90}, 85 (1993);
M. Tegmark, J. Silk, Blanchard, Astrophys. J. {\bf 420}, 484 (1993)
  (erratum {\bf 434}, 395);
A. Liddle, D.H. Lyth, MNRAS {\bf 273}, 1177 (1995), astro-ph/9409077
\bibitem{Primack}
J. Primack et al., Phys. Rev. Lett., 74, 2160 (1995)
\bibitem{BurTyt}
S. Burles, D. Tytler, submitted to Science, astro-ph/9603070
\end{thebibliography}
\end{document}